\documentclass[12pt]{article}
\usepackage{graphicx,amsmath,amssymb,booktabs,listings,url,subcaption,cancel}
\usepackage[a4paper,margin=3cm,bmargin=4.5cm]{geometry}
\usepackage{cite}
\usepackage{color}

\begin{document}

\def\a{\alpha}
\def\b{\beta}
\def\c{\varepsilon}
\def\d{\delta}
\def\e{\epsilon}
\def\f{\phi}
\def\g{\gamma}
\def\h{\theta}
\def\k{\kappa}
\def\l{\lambda}
\def\m{\mu}
\def\n{\nu}
\def\p{\psi}
\def\q{\partial}
\def\r{\rho}
\def\s{\sigma}
\def\t{\tau}
\def\u{\upsilon}
\def\v{\varphi}
\def\x{\xi}
\def\y{\eta}
\def\z{\zeta}
\def\D{\Delta}
\def\G{\Gamma}
\def\H{\Theta}
\def\L{\Lambda}
\def\F{\Phi}
\def\P{\Psi}
\def\S{\Sigma}

\def\w#1{_{\text{#1}}}
\def\GeV{\,\text{GeV}}
\newcommand\ewkino{{\tilde\chi^{0,\pm}_{\relax}}}
\newcommand\neut  [1][\relax]{{\tilde\chi^0_{#1}}}
\newcommand\charP [1][\relax]{{\tilde\chi^+_{#1}}}
\newcommand\charM [1][\relax]{{\tilde\chi^-_{#1}}}
\newcommand\charPM[1][\relax]{{\tilde\chi^\pm_{#1}}}
\newcommand\smuL{\tilde\mu\w L}
\newcommand\smuR{\tilde\mu\w R}
\newcommand\snumu{\tilde\nu_\mu}
\newcommand\stau{\tilde\tau}
\newcommand\mPT{\cancel{p}_{\text{T}}}
\def\TOFILL{{\color{red}\textbf{??????}}}

\def\o{\over}
\def\beq{\begin{eqnarray}}
\def\eeq{\end{eqnarray}}
\newcommand{\gsim}{ \mathop{}_{\textstyle \sim}^{\textstyle >} }
\newcommand{\lsim}{ \mathop{}_{\textstyle \sim}^{\textstyle <} }
\newcommand{\vev}[1]{ \left\langle {#1} \right\rangle }
\newcommand{\bra}[1]{ \langle {#1} | }
\newcommand{\ket}[1]{ | {#1} \rangle }
\newcommand{\EV}{ {\rm eV} }
\newcommand{\KEV}{ {\rm keV} }
\newcommand{\MEV}{ {\rm MeV} }
\newcommand{\GEV}{ {\rm GeV} }
\newcommand{\TEV}{ {\rm TeV} }
\def\Order{\mathop{\mathcal O}}
\def\sign{\mathop{\rm sign}}
\def\diag{\mathop{\rm diag}\nolimits}
\def\Spin{\mathop{\rm Spin}}
\def\SO{\mathop{\rm SO}}
\def\O{\mathop{\rm O}}
\def\SU{\mathop{\rm SU}}
\def\U{\mathop{\rm U}}
\def\Sp{\mathop{\rm Sp}}
\def\SL{\mathop{\rm SL}}
\def\tr{\mathop{\rm tr}}

\def\IJMP{Int.~J.~Mod.~Phys. }
\def\MPL{Mod.~Phys.~Lett. }
\def\NP{Nucl.~Phys. }
\def\PL{Phys.~Lett. }
\def\PR{Phys.~Rev. }
\def\PRL{Phys.~Rev.~Lett. }
\def\PTP{Prog.~Theor.~Phys. }
\def\ZP{Z.~Phys. }


\baselineskip 0.65cm

\begin{titlepage}

\begin{flushright}

\end{flushright}

\vskip 1.35cm
\begin{center}
\begingroup
\def\AFF#1{$^{({\it #1})}$}
{\large \bf Wino-Higgsino dark matter in MSSM from the $g-2$ anomaly}
\vskip 1.2cm
Sho Iwamoto\AFF{a}, Tsutomu T. Yanagida\AFF{b,c} and Norimi Yokozaki\AFF{d}
\vskip 1.4cm

\it \footnotesize
\AFF{a}ELTE E\"otv\"os Lor\'and University, P\'azm\'any P\'eter s\'et\'any 1/A, Budapest H-1117, Hungary\\[0.5em]
\AFF{b}Tsung-Dao Lee Institute, Shanghai Jiao Tong University, Shanghai 200240, China\\[0.5em]
\AFF{c}Kavli IPMU (WPI), UTIAS, The University of Tokyo, 5--1--5 Kashiwanoha, Kashiwa, Chiba 277--8583, Japan\\[0.5em]
\AFF{d}Zhejiang Institute of Modern Physics and Department of Physics, Zhejiang University, Hangzhou, Zhejiang 310027, China\\[0.5em]
\endgroup

\abstract{
In this letter, we show that the wino-Higgsino dark matter (DM) is detectable in near future DM direct detection experiments for almost all consistent parameter space in the spontaneously broken supergravity (SUGRA) if the muon $g-2$ anomaly is explained by the wino-Higgsino loop diagrams.
We also point out that the present and future LHC experiments can exclude or confirm this SUGRA explanation of the observed muon $g-2$ anomaly. 
}
\end{center}
\end{titlepage}
\renewcommand{\thefootnote}{\#\arabic{footnote}}

\setcounter{page}{2}

\noindent The new result on the muon $g-2$ measurement has been reported by the Fermilab group~\cite{g-2Seminar20210407}, which corresponds to a $4.2\sigma$-level discrepancy between the Standard Model prediction~\cite{Aoyama:2020ynm},
\begin{equation}
 \Delta a_{\mu}
 \equiv a_\mu^{\text{expm}} - a_\mu^{\text{SM}}
 =  \left(25.1\pm5.9\right) \times 10^{-10}.
\end{equation}
This new result is consisitent with the previous Brookhaven result~\cite{Bennett:2002jb,Bennett:2004pv,Bennett:2006fi}.

It is well known~\cite{Lopez:1993vi,Chattopadhyay:1995ae,Moroi:1995yh} that this anomaly can be easily explained in the minimal supersymmetric (SUSY) standard model (MSSM).
However, this solution potentially has a problem of the flavor-changing neutral currents (FCNC), since sleptons are predicted below 1 TeV.
To avoid too much FCNC, we assume the universality of the slepton masses at an ultraviolet (UV) scale such as the grand unified theory (GUT) scale or Planck scale.
In this case, there are two well-known parameter regions consistent with the $g-2$ anomaly within the MSSM:
one relies on the bino-slepton loop and the other utilizes the wino-Higgsino-slepton loop for the generation of the SUSY contribution $\Delta a_\mu^{\text{SUSY}}$ to the muon $g-2$.\footnote{See, e.g., Ref.~\cite{Endo:2017zrj} for other possibilities within the MSSM, such as those with $\sign(\mu)=-1$.}
In the former case, the bino can be the dominant dark matter (DM), whose detection probability in the direct detection experiments is very low and we can not expect the direct detection of the bino DM in near future \cite{Cox:2018vsv}.\footnote{ If we abandon the universality of the slepton masses, it is not the case \cite{Cox:2018qyi}.} 
In the latter case the, the DM is mixture of wino and Higgsino, i.e., ``wino-Higgsino DM,''  but its abundance is much smaller than the observed DM density \cite{Harigaya:2015kfa}.

\begin{figure*}[t]
  \centering
  \includegraphics[width=150pt]{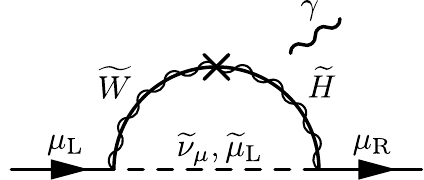}
  \caption{The dominant SUSY contribution to the muon $g-2$ in our scenario.}
\label{fig:diagram}
\end{figure*}

In this short letter, we consider the latter case, where the wino and Higgsino soft masses are $\Order(100)\GeV$ and $\Delta a_\mu^{\text{SUSY}}$ mainly comes from the diagram in Fig.~\ref{fig:diagram}, and show that we can detect such a subdominant wino-Higgsino DM in near future direct detection experiments.
We also discuss constraints and future prospects at collider experiments and provide benchmark points for LHC study.
Note that, in this letter, we consider an almost generic model of supergravity (SUGRA) under the assumption that the FCNC problem is solved with universal masses for squarks and sleptons. All the SUSY breaking parameters are given at the UV scale (GUT scale).
Because of this, our work is completely different from other related works such as Refs.\cite{Endo:2013bba, Cox:2018qyi, Endo:2020mqz, Chakraborti:2020vjp,Chakraborti:2021kkr}, where the soft SUSY breaking mass parameters are given at the low-energy scale and effects of renormalization group running are negligible.

The model we consider is based on the spontaneously broken SUGRA and has ten free parameters, $m_0$, $m_{H_u}^2$, $m_{H_d}^2$, $A_u$, $A_d(=A_e)$, $M_1$, $M_2$, $M_3$, $B_\mu/\mu$, and $\sign(\mu)$, after imposing the EWSB conditions.\footnote{To ensure the universal mass, $m_0$, it is assumed that the couplings between the matter fields and a SUSY breaking fields are identical, leading to $A_d=A_e$. The condition, $A_d=A_e$, is not important, though. }
Note that the FCNC problem is avoided because all the squarks and sleptons have the common soft mass, $m_0$, at the UV scale.\footnote{
As well-known examples to avoid the FCNC problem, a minimal Kahler potential or a sequestered Kahler potential~\cite{Inoue:1991rk,Randall:1998uk} (for the matter fields) leads to the common soft mass, $m_0$ or zero.  
}
We take $\sign(\mu)=+1$ to obtain $\Delta a_\mu^{\text{SUSY}}>0$.
Furthermore, since the effects of $m_0$ can almost be absorbed into the bino mass $M_1$, we can take $m_0=0$ without a significant change in our conclusion.
We also checked that the $A_d$ dependence of the DM detection is very weak, for which we assume $A_d=0$ at the GUT scale.
Accordingly, we have seven parameters defined at the GUT scale,
$M_1, M_2, M_3, A_u, m_{H_u}^2, m_{H_d}^2$ and $B_\mu/\mu$.\footnote{With the choice of the model parameters, 
 this SUGRA model is almost equivalent to a gaugino mediation model in Ref.~\cite{Harigaya:2015kfa}, where the SUSY CP and flavor problems are solved, and the origin of the non-universal gaugino masses is naturally explained.}
They are equivalent to the following parameter, which we choose as the model parameters in this work:
\begin{equation}
 M_1, M_2, M_3, A_u, \mu, m_A\text{, and~}\tan\beta,
\end{equation}
where $M_{1,2,3}$ ($A_u$) are the gaugino soft mass (scalar cubic coupling) defined at the GUT scale, $\mu$ is the Higgsino invariant mass at the low-energy scale (stop mass scale).
$m_A$ is the pole mass of the CP-odd Higgs boson,
 and $\tan\beta$ is the ratio of Higgs vacuum expectation values.

Our SUGRA model, in which $\Delta a_\mu^{\text{SUSY}}$ mainly comes from the wino-Higgsino loops (Fig.~\ref{fig:diagram}) and the DM is wino-Higgsino mixture, is characterized by sub-TeV $\mu$ and $M_2$;
Fig.~2--3 shows the consistent parameter region on the $\mu$--$M_2$ plane.
The other parameters are chosen to be $(M_1, M_3, A_u, m_A)$ = $(2100, 2500, -1000, 2500)$\,GeV and $\tan\beta=24$.
The SUSY mass spectra are calculated by \texttt{SuSpect 2.43}~\cite{Djouadi:2002ze} with modifications: the input scale of the model parameters (i.e. the GUT scale) is fixed to be $10^{16}$\,GeV and iteration procedures are slightly changed (see Appendix).
The Higgs boson mass and $\Delta a_\mu^{\text{SUSY}}$ are computed by \texttt{FeynHiggs 2.18.0}~\cite{Heinemeyer:1998yj,Heinemeyer:1998np,Degrassi:2002fi,Frank:2006yh,Hahn:2013ria,Bahl:2016brp,Bahl:2017aev,Bahl:2018qog}. For the estimation of $\Delta a_\mu^{\text{SUSY}}$, we include $\tan\beta$-enhanced corrections at the two-loop level~\cite{Marchetti:2008hw}.
We also use \texttt{MicrOMEGAs 5.2.7.a}~\cite{Belanger:2001fz,Belanger:2004yn} to obtain the relic density of the LSP, $\Omega\w{LSP}$, and its spin-independent cross section to nucleons, $\sigma\w{SI}$.

\begin{figure*}[t]
  \centering
  \includegraphics[scale=0.8,trim=70 0 100 20]{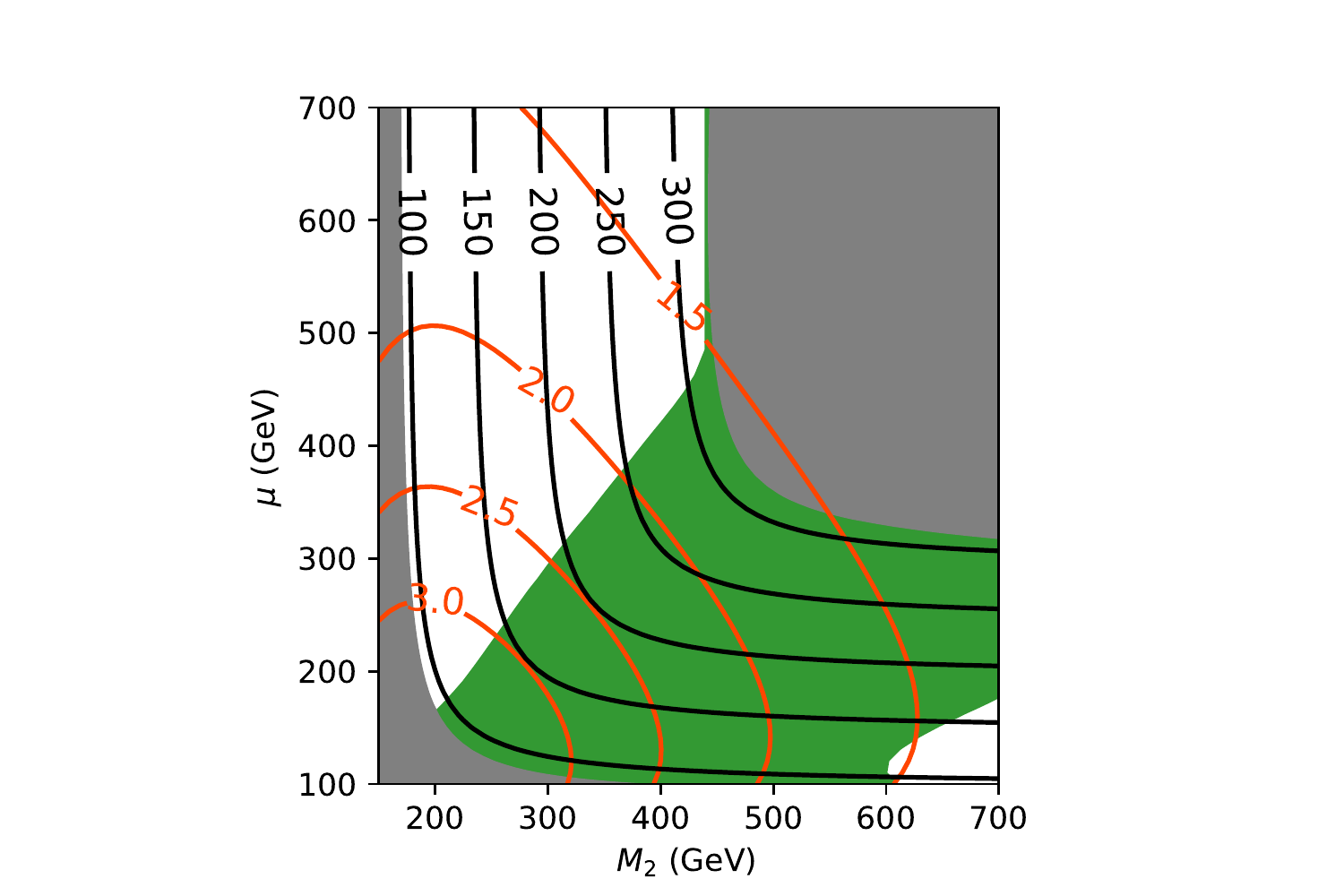}
  \caption{\label{fig:1}%
Summary of our benchmark parameter space, where
the red contours show the SUSY contribution to the muon $g-2$ $(a_\mu^{\text{SUSY}}\times10^9)$ and
the lightest neutralino mass is displayed by the black contours ($m_{\tilde\chi_1^0}/\text{GeV}$).
See text for details.
}
\end{figure*}

\begin{figure}[t]
 \centering
\begin{subfigure}[b]{0.49\textwidth}
  {\includegraphics[scale=0.8,trim=70 0 100 20]{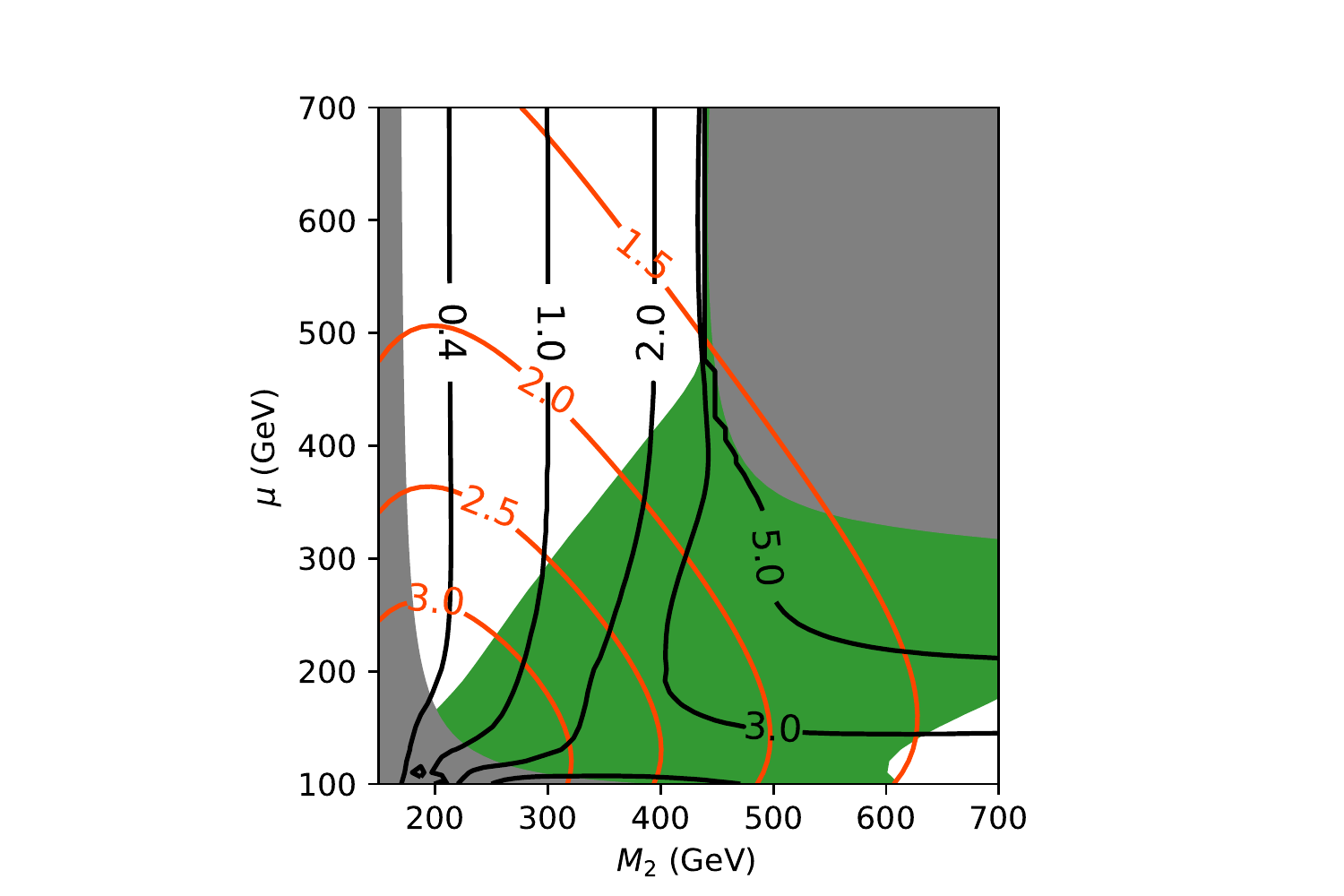}}
 \caption{$\Omega\w{LSP}/\Omega_{\rm CDM}$~[\%]}
\end{subfigure}
\begin{subfigure}[b]{0.49\textwidth}
  {\includegraphics[scale=0.8,trim=70 0 100 20]{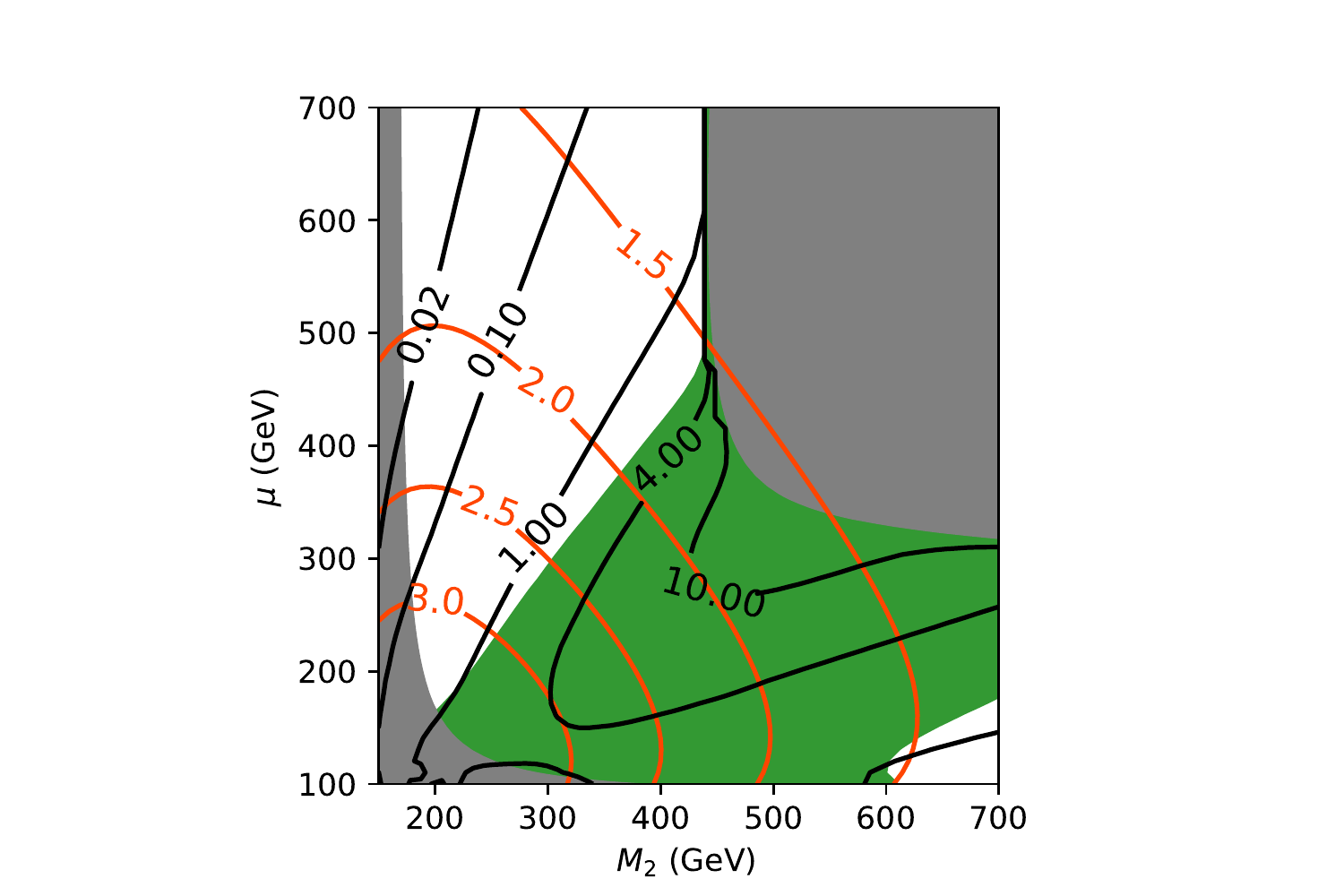}}
 \caption{$\sigma\w{SI}^{\text{eff}}/10^{-10}$ pb}
\end{subfigure}
  \caption{%
The same as Fig.~\ref{fig:1}, but the black contours in the left figure show the relic abundance of the lightest neutralino $\tilde\chi^0_1$ normalized by the total DM density in percentage terms, while in the right figure they show the effective spin-independent cross section for the DM direct detection is given in units of $10^{-10}\text{\,pb}=10^{-46}\text{\,cm}^2$.
}
\label{fig:2}
\end{figure}

\begin{figure*}[t]
  \centering
  \includegraphics[scale=0.8,trim=70 0 100 20]{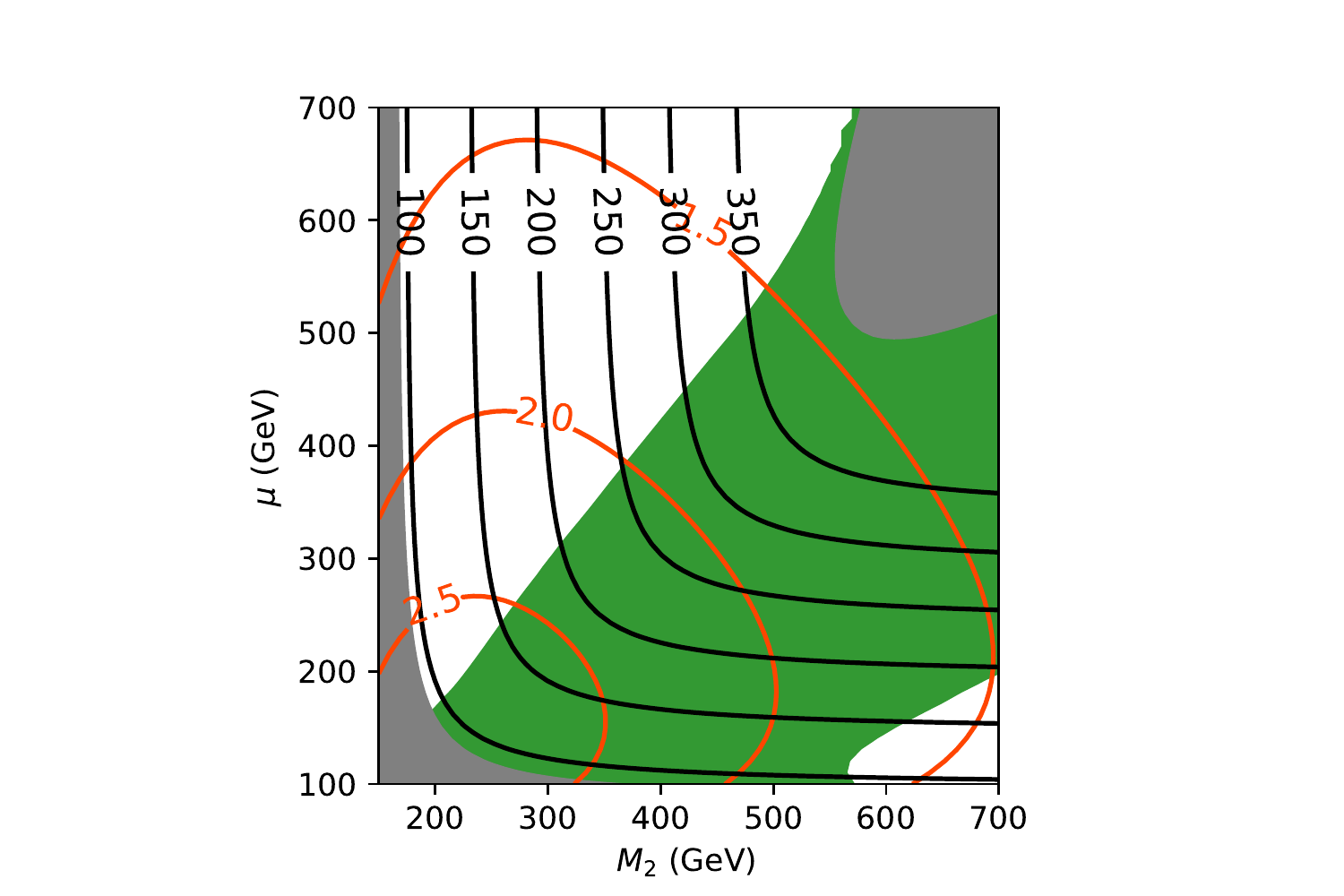}
  \caption{\label{fig:3}%
The same as Fig.~\ref{fig:1}, but the parameters are taken to be $(M_1, M_3, A_u, m_A)=(3800,2500,-1000,2000)\GeV$ and $\tan\beta=40$.
}
\end{figure*}

\begin{figure}[t]
 \centering
\begin{subfigure}[b]{0.49\textwidth}
  {\includegraphics[scale=0.8,trim=70 0 100 20]{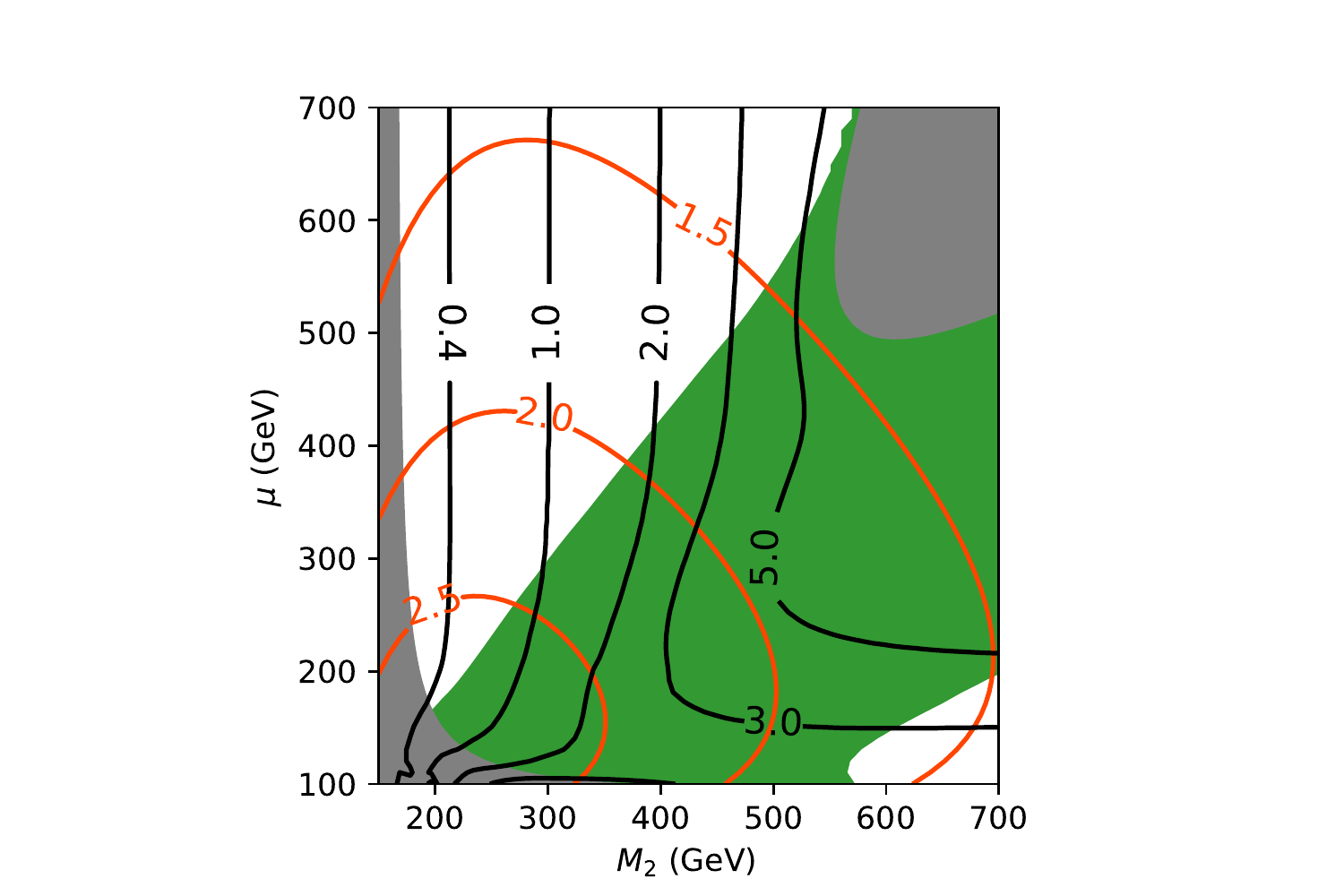}}
 \caption{$\Omega\w{LSP}/\Omega_{\rm CDM}$~[\%]}
\end{subfigure}
\begin{subfigure}[b]{0.49\textwidth}
  {\includegraphics[scale=0.8,trim=70 0 100 20]{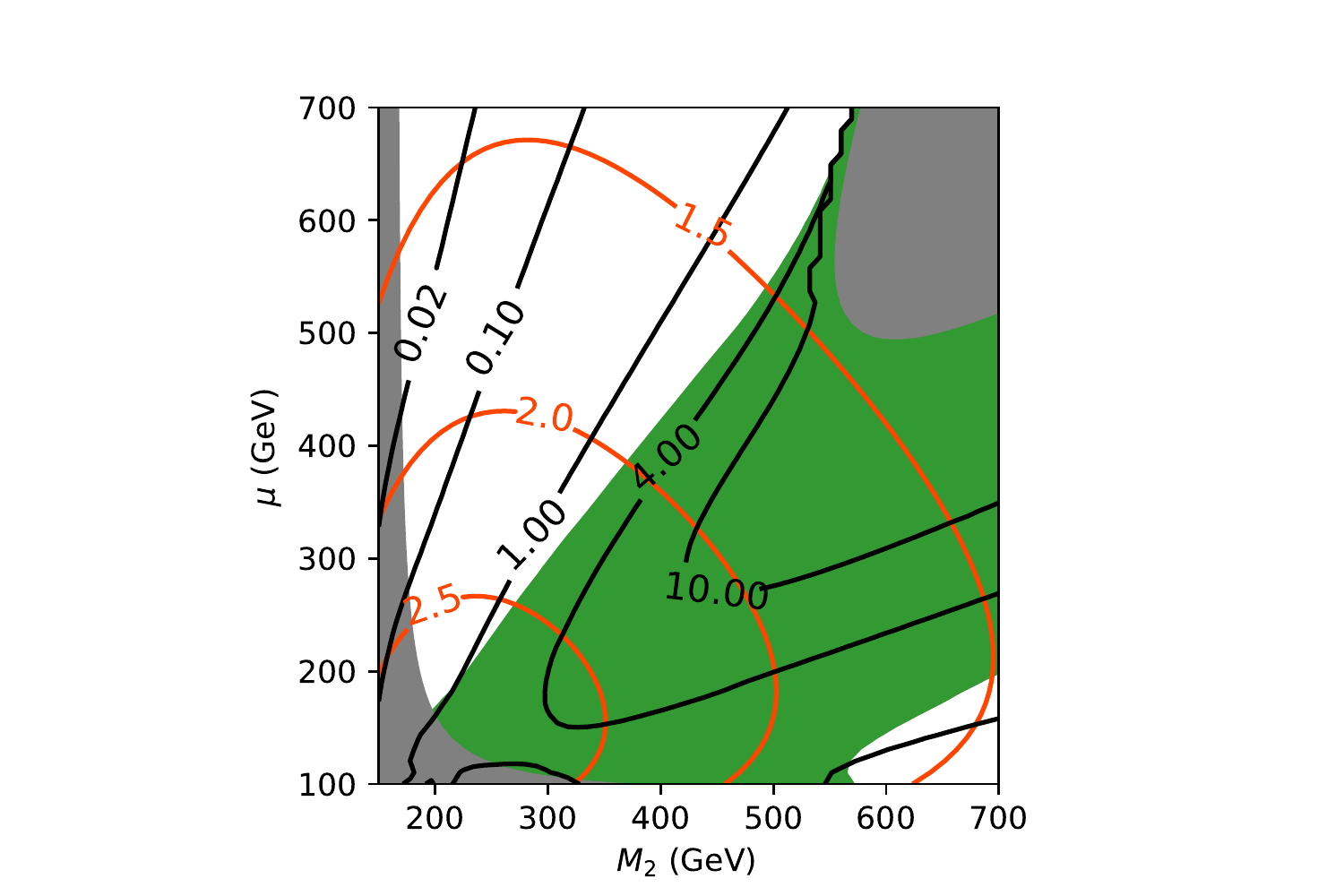}}
 \caption{$\sigma\w{SI}^{\text{eff}}/10^{-10}$ pb}
\end{subfigure}
  \caption{%
The same as Fig.~\ref{fig:2}, but the parameters are taken to be $(M_1, M_3, A_u, m_A)=(3800,2500,-1000,2000)\GeV$ and $\tan\beta=40$.
}
\label{fig:4}
\end{figure}

\begin{table}[t]
\centering
 \caption{\label{tab:BP}Benchmark points for the muon $g-2$ anomaly.
The other UV parameters are fixed to $(M_1, M_3, A_u, m_A)=(2100,2500,-1000,2500)\GeV$ and $\tan\beta=24$ for BP-WH1 to BP-WH4, and $(M_1, M_3, A_u, m_A)=(3800,2500,-1000,2000)\GeV$ and $\tan\beta=40$ for BP-WH5.
All the colored SUSY particles are heavier than 3 TeV and thus not collected here.
Because $M_1\gg M_2$, the heaviest neutralino $\tilde\chi^0_4$ is bino-like, while the other neutralinos are wino-Higgsino mixture.
Lifetime is calculated by \texttt{SOFTSUSY 4.1.10}~\cite{Allanach:2001kg,Allanach:2017hcf} and LHC production cross sections are by \texttt{Prospino 2.1}~\cite{Beenakker:1999xh}.
}
\vspace{1em}
\begin{minipage}{\textwidth}\renewcommand{\arraystretch}{1.1}
\begin{tabular}{cccccc}\toprule
& BP-WH1 & BP-WH2 & BP-WH3 &  BP-WH4\footnote{Excluded by XENON1T 
experiment but prepared for collider studies.} &BP-WH5 \\\midrule
$M_2$/GeV                   & 200 & 300 & 250 &  500 & 250\\
$\mu$/GeV                   & 300 & 450 & 250 &  200 & 250 \\
$m(\neut[1],\charPM[1])$/GeV& (112, 112) & (201, 201) & (143, 145)  & (188, 193) & (146, 147)\\
$m(\neut[2])$/GeV           & 314        & 464        & 263         
      & 211     & 263   \\
$m(\neut[3],\charPM[2])$/GeV& (322, 330) & (470, 476) & (284, 290) &  (402, 402) & (284, 290)\\
$m(\neut[4])$/GeV           & 927        & 927        & 927              & 927       & 1457 \\
$m(\smuL,\smuR)$/GeV        & (459, 615) & (480, 615) & (469, 614)  & (542, 609) & (736, 1314)\\
$m(\stau_1,\stau_2)$/GeV    & (286, 354) & (303, 365) & (298, 346)  & (321, 407) & (292, 899)\\
\midrule
$m_h$/GeV                                              & 124.0 & 123.5 & 123.6  & 123.6 & 124.2\\
$\Delta a_\mu^{\text{SUSY}}\times10^{10}$              & 28.0 & 19.9 & 29.2  & 19.3 & 25.5\\
$\Omega\w{LSP}/\Omega\w{CDM}$                          & 0.0033 & 0.010 & 0.006  & 0.040 & 0.007\\
$\sigma\w{SI}^{\rm eff}/10^{-10}$ pb                   & 0.14 & 0.3 & 0.9  & 4.3 & 1.0 \\
\midrule
$\tau(\tilde\chi^\pm_1)$/s                             & $4\times10^{-12}$ & $2\times10^{-11}$ & $8\times10^{-14}$ &  $7\times10^{-17}$ & $8\times10^{-14}$\\
$\mathop{\text{Br}}(\tilde\mu\w L\to\charM[1]\nu)$     & 0.63    & 0.66   
 & 0.55      & 0.091\footnote{$\mathop{\text{Br}}(\tilde\mu\w L\to\charM[2]\nu)=0.56$.} & 0.50\\
$\sigma(pp\to\neut[i]\charPM[2])_{\text{13\,TeV}}${}\footnote{For BP-WH1 to WH3 and WH5, $i$ is summed over 2--3, while BP-WH4 only includes $i=3$.}
                                                       & 144\,fb &  30\,fb & 263\,fb & 102\,fb & 260\,fb\\
\bottomrule
\end{tabular}
\end{minipage}
\end{table}

The results are summarized in Fig.~\ref{fig:1}, where $ \Delta a_\mu^{\text{SUSY}}$ and the LSP mass are respectively shown by the red and black contours.
The upper-right gray-shaded regions are excluded due to the stau LSP, while the left gray-shaded region is excluded by the LEP2 experiment~\cite{LEP2SUSYWG:01-03.1}.
The green-shaded region is excluded by the XENON1T experiment on DM direct detection.
The remaining region is motivated by the muon $g-2$ anomaly and we will discuss constraints on this benchmark space from DM direct detection and from LHC experiments.

Because of  $M_1\gg M_2\sim\mu$, the lightest neutralino, being the LSP, is the mixture of wino and Higgsino and can be a subdominant component of the DM.
Its relic density is shown by the black contours Fig.~\ref{fig:2}(a) in percentage terms, where the values are normalized by the total DM abundance, $\Omega_{\rm CDM}h^2\simeq0.12$.
Although it is only a few percent of the total DM, their spin-independent cross section to protons are sizable.
As shown in Fig.~\ref{fig:2}(b), the effective spin-independent cross section of the DM direct detection, defined by $\sigma_{\rm SI}^{\rm eff} \equiv\Omega\w{LSP}/\Omega\w{CDM}\times\sigma\w{SI}$, is as large as $10^{-46}\,\text{cm}^2$ and the green-shaded region is excluded by the XENON1T experiment~\cite{Aprile:2018dbl}.
Most of the remaining region will be probed by future direct-detection experiments, such as  DarkSide-20k~\cite{Aalseth:2017fik}, LZ~\cite{Akerib:2018lyp}, PandaX-4T~\cite{Zhang:2018xdp}, and XENONnT~\cite{Aprile:2020vtw}.

Figures \ref{fig:3}--\ref{fig:4} show the results (the LSP mass, the relic abundance and the SI cross section) for $(M_1, M_3, A_u, m_A)=(3800,2500,-1000,2000)\GeV$ and $\tan\beta=40$. The results are very similar to those shown in Fig.~\ref{fig:1}--\ref{fig:2}. The main difference is that the left-handed selectron and smuon can be as heavy as 740-760 GeV in the region consistent with the muon $g-2$ at 1$\sigma$ level.

The LHC phenomenology of our scenario is rather complicated because of compressed spectra and the mixed nature of neutralinos.
Therefore, we do not perform full analyses in this work, but instead introduce a few benchmark points (BPs) for future analyses and provide general brief discussion on the LHC phenomenology.
The BPs are shown in Table~\ref{tab:BP}; note that BP-WH4 is excluded by XENON1T as discussed above, but introduced for completeness.
Colored SUSY particles are not displayed because they are heavy due to large $M_3$ and beyond the present LHC limits.
We thus focus on non-colored SUSY particle production at LHC, which are of our interest because our scenario is motivated by the muon $g-2$ anomaly and wino-Higgsino partial dark matter.

Chargino searches based on disappearing-track signature are promising for the wino DM.
Indeed, $\charPM[1]$ is almost wino-like in the upper-left corner of Fig.~2 and its lifetime, $\tau(\charPM[1])$, is longer than $10^{-11}$s; a portion of that region is thus expected to be excluded by the results in Refs.~\cite{Sirunyan:2020pjd,ATLAS:2021ttq}.
However, those limits do not cover the region with $M_2\simeq\mu$ because, due to the Higgsino component of $\charPM[1]$, the lifetime is shorter and the production cross section is smaller compared to wino-like chargino.

Although sleptons are as light as $\sim400\GeV$, the slepton channel $pp\to\tilde l\tilde l^*\to2l+\mPT$~\cite{Aad:2019vnb,Sirunyan:2020eab} does not provide significant limit, either.
This is mainly because, unlike bino-LSP models studied in Refs.~\cite{Endo:2020mqz,Endo:2021zal}, the sleptons are allowed to decay into $\charPM[1]$ without emitting any hard charged leptons, as depicted in Table~\ref{tab:BP}.

The region with $M_2\simeq\mu$ are therefore to be searched for by the production of heavier electroweakinos, $\tilde\chi^\pm_2$ and $\tilde\chi^0_{2,3}$.
In fact, as examined in Refs.~\cite{Endo:2020mqz}, SUSY models with sizable $\Delta a_\mu^{\text{SUSY}}$ from wino-Higgsino loop (Fig.~\ref{fig:diagram}) are typically searched for by electroweakino pair-production, which yields signatures with two SM bosons ($W^\pm$, $h$, $Z$) with large missing transverse momentum.
Unlike bino-LSP models studied in Refs.~\cite{Endo:2013bba,Endo:2020mqz,Endo:2021zal}, the expected signature is diverse and involved because various production channels are expected and the electroweakinos are allowed to decay into all the SM bosons.
Dedicated LHC analyses are called for and, thanks to the sizable production cross section, they are expected to probe the whole parameter space that are motivated by the muon $g-2$ anomaly.

\subsubsection*{Acknowledgments}
T.~T.~Y.~is supported in part by the China Grant for Talent Scientific Start-Up Project and the JSPS Grant-in-Aid for Scientific Research No.\,17H02878, and No.\,19H05810, and by World Premier International Research Center Initiative (WPI Initiative), MEXT, Japan.

\lstset{basicstyle=\ttfamily\footnotesize, commentstyle=\ttfamily}
\subsubsection*{Appendix: SLHA Input}
Our benchmark parameter space corresponds to the following SLHA input for spectrum generators such as \texttt{SuSpect}~\cite{Djouadi:2002ze} or \texttt{SOFTSUSY}~\cite{Allanach:2001kg}\footnote{``Number of loops in Higgs mass computation'' should be set to 2.}:
\begin{lstlisting}
BLOCK MODSEL
     1            1         # mSUGRA 
BLOCK MINPAR
     1     0.00000000E+00   # m0
     2     1.00000000E+03   # M1/2
     3     2.40000000E+01   # tan(beta)
     4     1.00000000E+00   # sign(mu)
     5     0.00000000E+00   # A
BLOCK EXTPAR
     1     2.10000000E+03   # M1
     2        (varied)      # M2
     3     2.50000000E+03   # M3
    11    -1.00000000E+03   # At
    12     0.00000000E+00   # Ab
    13     0.00000000E+00   # Atau
    23        (varied)      # mu
    26     2.50000000E+03   # mA0
\end{lstlisting}
To make the SUSY spectrum convergent during iteration processes, one may need to modify the spectrum generators.
For instance, if one uses \texttt{SuSpect\,2.52}, one needs to take $m_{H_u}=m_{H_d}=0$ at the GUT scale for first few iterations; specifically, \texttt{suspect2.52.f} should be modified by the following patch:
\begin{lstlisting}[frame=single]
@@ -231,4 +231,5 @@
 c  reinitialize various control parameters + other parameters:
+           myiter=0
            do ierr=1,10
            errmess(ierr)=0.d0
            enddo
@@ -1099,2 +1100,7 @@
        ifirst=0
        jfirst=0
+       if(myiter.lt.4) then
+         y(12)=0.
+         y(13)=0.
+       endif
+       myiter=myiter+1
\end{lstlisting}

\bibliographystyle{utphys28mod}
\bibliography{g-2_sugra}
\end{document}